\documentclass{emulateapj}
\slugcomment{} 
\shorttitle{Gravitational Lens Time Delays}
\shortauthors{Oguri}
\begin{document}
\title{Gravitational Lens Time Delays: A Statistical Assessment of
  Lens Model Dependences and Implications for the Global Hubble
  Constant}  
%
\author{
Masamune Oguri
}
\affil{Kavli Institute for Particle Astrophysics and
Cosmology, Stanford University, 2575 Sand Hill Road, Menlo Park, 
CA 94025.}
\email{oguri@slac.stanford.edu} 
%
%
\begin{abstract}
Time delays between lensed multiple images have been known to provide an
interesting probe of the Hubble constant, but such application is often
limited by degeneracies with the shape of lens potentials. We propose a
new statistical approach to examine the dependence of time delays on
the complexity of lens potentials, such as higher-order perturbations,
non-isothermality, and substructures. Specifically, we introduce a
reduced time delay of the dimensionless form, and explore its behavior
analytically and numerically as a function of the image configuration
that is characterized  by the asymmetry and opening angle of the image
pair. In particular we derive a realistic conditional probability
distribution for a given image configuration from Monte-Carlo
simulations. We find that the probability distribution is sensitive to
the image configuration such that more symmetric and/or smaller opening
angle image pairs are more easily affected by perturbations on the primary
lens potential. On average time delays of double lenses are less
scattered than those of quadruple lenses. Furthermore, the realistic
conditional distribution allows a new statistical method to constrain
the Hubble constant from observed time delays. We find that 16 published
time delay quasars constrain the Hubble constant to be $H_0=70\pm 6
\,{\rm km\,s^{-1}Mpc^{-1}}$, where the value and its error are
estimated using jackknife resampling. Systematic errors coming
from the heterogeneous nature of the quasar sample and the uncertainty 
of the input distribution of lens potentials can be larger than the
statistical error. After including rough estimates of the sizes of
important systematic errors, we find $H_0=68\pm 6({\rm stat.})\pm8({\rm 
 syst.}) \,{\rm km\,s^{-1}Mpc^{-1}}$. The reasonable agreement of the
value of the Hubble constant with other estimates indicates the
usefulness of our new approach as a cosmological and astrophysical
probe, particularly in the era of large-scale synoptic surveys.    
\end{abstract}


\section{Introduction}\label{sec:intro}

It has been known that time delays between multiple images of strong
gravitational lens systems offer an interesting method to measure the
Hubble constant $H_0$, the most fundamental cosmological parameter
that governs the length and time scale of our universe
\citep{refsdal64}. A huge advantage of this method is that it does not
rely on so-called distance ladder and can measure the global Hubble
constant independently of any local measurements. Motivated by this
time delays have been measured in more than 10 lensed quasar systems
\citep[see, e.g.,][]{kochanek06a}. 
The situation as it presents is, however, somewhat confusing and
controversial. \citet{kochanek02,kochanek03} claimed from the analysis
of several lens systems that the Hubble constant should be relatively
low, $H_0\sim 50 \,{\rm km\,s^{-1}Mpc^{-1}}$. The time delay of SDSS
J1650+4251 also prefers the low Hubble constant \citep{vuissoz07}.
On the other hand, \citet{koopmans03} performed systematic mass
modeling of B1608+656 using all available data from radio to optical
and found constrained the value of the Hubble constant to be
$H_0=75^{+7}_{-6}\,{\rm km\,s^{-1}Mpc^{-1}}$. The analysis of the
smallest separation lens B0218+357 yields $H_0=78\pm 3\,{\rm
  km\,s^{-1}Mpc^{-1}}$ \citep{wucknitz04}. By combining time delays
in 10 lensed quasar systems \citet{saha06a} obtained
$H_0=72^{+8}_{-11}\,{\rm km\,s^{-1}Mpc^{-1}}$.
 
The large variation of derived values of the Hubble constant from time
delays reflects the fact that time delays are quite sensitive to
mass distributions of lens objects, which leads to strong
degeneracies between lens mass profiles and the Hubble constant.
The most fundamental, mathematically rigorous degeneracy is the
mass-sheet degeneracy \citep{falco85}; inserting an uniform sheet
instead of decreasing the mass normalization of lenses changes
estimated values of the Hubble constant while leaving unchanged the
image observables. This degeneracy implies that the derived values of
the Hubble constant also degenerates with radial density profiles of
of lens galaxies \citep{refsdal94,witt95,keeton97b,witt00,saha00,tada00,
wucknitz02,treu02,kochanek02,kochanek03,oguri03b,rusin03,schechetr05,
mortsell05,kawano06}. In addition, it may degenerate with the angular
structure of lenses as well \citep{zhao03,saha06b}. The sensitivity of
time delays on mass profiles suggests that by assuming the Hubble
constant, which can be determined by other methods
\citep{freedman01,tegmark06,spergel07}, we can put constraints on mass
distributions of lenses, in particular radial density profiles
\citep[e.g.,][]{oguri04a,kochanek06b,dobke06}.  

One way to overcome the degeneracies is to adopt lens systems 
whose lens potentials can well be constrained by many observational
constraints.  An example of such constraints comes from lensed images of
quasar host galaxies. The lensed host galaxies often form Einstein
rings, which accurately and independently determines the structure of
the lens potential as well as the shape of the lensed host galaxy
\citep[e.g.,][]{keeton00,kochanek01,koopmans03}. The small-scale
structure of lensed quasars, such as radio jets and sub-components,
determine lens potentials in a similar way as Einstein rings 
\citep[e.g.,][]{cohn01,wucknitz04}.  Furthermore, the measurement of
velocity dispersions of lens galaxies serves as useful constraints on
mass distributions, helping to break degeneracies between mass models
\citep[e.g.,][]{treu02}. Lens systems for which these strong
additional constraints are available, sometime referred as ``golden
lenses'', have been thought to be an effective probe of the Hubble
constant. 

Another potentially powerful, but less studied, method to measure the
Hubble constant is a statistical approach. Even if each individual
lens lacks strong constraints that allow detailed
investigations of the mass distribution, by combining many lens
systems one can put tight constraints on the Hubble constant. 
As mentioned above, this approach was in some sense demonstrated
recently by \citet{saha06a} who combined 10 lensed quasar systems to
constrain the Hubble constant. A caveat is that lensed quasars
sometimes suffer from selection effects. For instance, brighter lensed
quasars  with larger image separations are more likely to lie in dense
environments such as groups and clusters \citep{oguri05b,oguri06a},
thus the Hubble constant from those lensed quasars may be
systematically higher than the true value without any correction of
the effect of dark matter along the line of sight. This indicates the
importance of well-defined statistical sample of lensed quasars for
which we can quantitatively estimate and correct the selection
effect. While the statistical sample has been very limited so far,
containing $\sim 20$ lenses even for the largest lens sample
\citep{myers03,browne03}, larger lens surveys will be soon available
by ongoing/future lens surveys such as done by the Sloan Digital Sky
Survey \citep{oguri06b}, the Large Synoptic Survey
Telescope (LSST)\footnote{http://www.lsst.org/}, and the
Supernova/Acceleration Probe
(SNAP)\footnote{http://snap.lbl.gov/}. Future lens surveys will also
find strong lensing of distant supernovae \citep[e.g.,][]{oguri03a} for
which time delays can easily be measured \citep[but
  see][]{dobler06}. Therefore the statistical approach is growing its
importance.    
   
In this paper, we study how time delays depend on various properties
of the lens potential and image configurations, which is essential for 
the determination of the Hubble constant from time delays. Using both
analytic and numerical methods, we show how time delays are affected
by several complexity of the lens potential, such as radial mass
profiles, external perturbations, higher order multipoles, and
substructures. We then derive the expected distributions of time
delays by adopting realistic lens potentials. The distribution, in
turn, can be used to derive statistically the value of the Hubble
constant from observed time delays. This approach differs from the
statistical argument by \citet{saha06a} in the sense that they fit
image positions of individual lens systems to constrain the Hubble
constant and then combined results of all the lens systems: Our new
approach does not even require modeling of each lens system. This has an
advantage that we can include lens systems that have too few constraints
to determine the lens potential. In this sense the approach is extension
of study by \citet{oguri02} in which only spherical  halos are
considered to compute time delay distributions. We note that the
methodology is similar to that adopted by \citet{keeton03a,keeton05} who
derived distributions of flux ratios of image pairs to identify
small-scale structure in lens galaxies.  

In addition to the measurement of the Hubble constant, the sensitivity
of time delays on mass models, which we explore in this paper, offers 
guidance on the usefulness of each time delay measurement as a
cosmological or astrophysical probe: If time delays at some image
configuration is quite sensitive to detailed structure of the lens
such as higher-order multipole terms with small amplitudes or
substructures, which are difficult to be constrained even for
best-studied lens systems,  it is almost hopeless to use these time
delays to extract either radial mass profiles or the Hubble
constant. Our result can be used to assess quantitatively which lens
systems are more suitable for detailed studies, i.e., less sensitive to
the complexity of the lens potential. There has been several insightful
analytic work \citep[e.g.,][]{witt00,kochanek02}, but here we perform
more systematic and comprehensive survey of model dependences of time
delays by parameterizing image configurations of lensed quasar systems
using dimensionless quantities.  

This paper is organized as follows. In \S \ref{sec:model} we introduce
several dimensionless quantities that are used to explore the model
dependence of time delays. We study the sensitivity of time delays on
the various lens potentials analytically in \S \ref{sec:ana}. Section
\ref{sec:pot} is devoted to construct the conditional distribution of
time delays from realistic Monte-Carlo simulations. We compare it with
observed time delays in \S \ref{sec:obs}, and constrain the Hubble
constant in \S \ref{sec:hub}. Finally discussion of our results and
conclusion are given in \S \ref{sec:dis}. Throughout the paper we
adopt a flat universe with the matter density $\Omega_M=0.24$ and the
cosmological constant $\Omega_\Lambda=0.76$ \citep{tegmark06}, although
our results are only weakly dependent on the specific choice of the
cosmological parameters.  The Hubble constant is sometimes described
by the dimensionless form $h\equiv H_0/(100\,{\rm km\,s^{-1}Mpc^{-1}})$.

\section{Characterizing Time Delay Quasars}\label{sec:model}

Let us consider a lens system in which a source at ${\mathbf u}=(u,v)$
is multiply imaged at the image positions ${\mathbf x}_i=(x_i,y_i)$.
We also use the polar coordinates for the image positions, ${\mathbf
  x}_i=(x_i,y_i)=(r_i\cos\theta_i,r_i\sin\theta_i)$. 
We always choose the center of the lens object as the origin of the
coordinates. Time delays between these multiply images are given by
\citep[e.g.,][]{blandford86}
\begin{eqnarray}
 \Delta
  t_{ij}&=&\frac{1+z_l}{2c}\frac{D_{\rm ol}D_{\rm os}}{D_{\rm ls}}\nonumber\\
&&\hspace*{-5mm}\times\left[({\mathbf x}_i-{\mathbf u})^2-({\mathbf x}_j-{\mathbf u})^2
 -2\phi({\mathbf x}_i)+2\phi({\mathbf x}_j)\right],
\label{eq:td}
\end{eqnarray}
where $z_l$ is the redshift of the lens, $c$ is the speed of light, and
  $D_{\rm ol}$, $D_{\rm os}$, and $D_{\rm ls}$ are angular diameter
  distances from the observer to the lens, from the observer to the
  source, and from the lens to the source, respectively. The lens
  potential $\phi({\mathbf x})$ is related to the surface mass
  density of the lens $\Sigma({\mathbf x})$ by the Poisson equation:
\begin{equation}
 \nabla^2\phi({\mathbf x})=2\kappa({\mathbf x})
=2\frac{\Sigma({\mathbf x})}{\Sigma_{\rm crit}},
\end{equation}
with $\Sigma_{\rm crit}=c^2D_{\rm os}/(4\pi G D_{\rm ol}D_{\rm ls})$
being the critical surface mass density ($G$ is the gravitational
constant). Note that image positions and source positions are
related by the lens equation
\begin{equation}
 {\mathbf u}={\mathbf x}_i-\nabla\phi({\mathbf x}_i).
\end{equation}

Equation (\ref{eq:td}) involves unobservable quantities such as
${\mathbf u}$ and $\phi({\mathbf x})$, indicating that time delays in
general depend on details of mass models. However, \citet{witt00} has
shown that for generalized isothermal potential 
\begin{equation}
 \phi({\mathbf x})=r F(\theta),
\label{eq:iso}
\end{equation}
where $F(\theta)$ is an arbitrary function of $\theta$, time delays
can be expressed in a very simple form involving only the 
observed image positions: 
\begin{equation}
 \Delta t_{ij}=\frac{1+z_l}{2c}
\frac{D_{\rm ol}D_{\rm os}}{D_{\rm ls}}(r_j^2-r_i^2),
\end{equation}
where $r_i$ denote the distance of image $i$ from the center of the lens
galaxy. Motivated by this analytic result, in this paper we consider
the {\it reduced time delay} that is defined by:
\begin{eqnarray}
 \Xi&\equiv& \left|\frac{\Delta t_{ij}}{r_j^2-r_i^2}
\right|\frac{2c}{1+z_l}\frac{D_{\rm ls}}{D_{\rm ol}D_{\rm os}}\nonumber\\
&=&\left|\frac{({\mathbf x}_i-{\mathbf u})^2-
({\mathbf x}_j-{\mathbf u})^2-2\phi({\mathbf x}_i)
+2\phi({\mathbf x}_j)}{r_j^2-r_i^2}\right|.
\label{eq:xi_def}
\end{eqnarray}
In this definition, the reduced time delay $\Xi$ is always unity if the
lens potential can be expressed by equation (\ref{eq:iso}), but can
deviate from unity if the lens potential has more complicated
structures. In particular, from the analysis of \citet{kochanek02} we
obtain $\Xi=2(1-\langle\kappa\rangle)$ at the lowest order of expansion,
where $\langle\kappa\rangle$ is the average surface density in the
annulus bounded by the images. This indicates that the deviation from
the isothermal mass distribution has a direct impact on the reduced
time delay. But as we will see later it is affected to some extent by
other factors such as external perturbations or small-scale
structures as well. Thus we can regard the reduced time delay as a
measure of the complexity of the lens. In addition, equation
(\ref{eq:xi_def}) indicates that we can compute reduced time delays
for observed lensed quasar systems with measured time delays by
assuming the value of the Hubble constant. In this sense, the reduced
time delay $\Xi$ is a key quantity that represents a link between
measured time delays of lens systems and theoretical lens models. We
point out that $\Xi$ is dimensionless because time delays are
proportional to the square of the size of a lens (the Einstein
radius). This allows us to directly compare values of reduced time
delays for different lens systems that have different sizes.  

We note that \citet{saha04} adopted similar but different
dimensionless quantity to explore the dependence of time delays on
mass models.  In the paper, the parameter essentially same as equation
(\ref{eq:xi_def}) was also considered, but it was discarded because of
the correlation with time delays. However, the correlation just
reflects the effect of surrounding dark matter that is larger for
wider separation lenses \citep{oguri05b,oguri06a}. Put another way,
such correlation is naturally expected from very different mass
distributions and environments between small ($\sim 1''$) and large
($\gtrsim 3''$) separation lenses. Indeed the apparent lack of the
correlation between time delays and the scaled time delays in
\citet{saha04} comes mostly from the large scatter among different
image configurations: The effect of external mass is hindered by the
large scatter his parametrization involves.  Therefore in this paper
we propose equation  (\ref{eq:xi_def}) as useful quantity to extract
the mass dependences on time delays. 

Previous analytic calculations of time delays suggest that the model
dependence of time delays is a strong function of image configurations
\citep{witt00,kochanek02}. In this paper we characterize image
configurations by the following two parameters. One is the {\it
  asymmetry} of the images define by
\begin{equation}
 R_{ij}\equiv \left|\frac{r_j-r_i}{r_j+r_i}\right|.
\label{eq:def_rij}
\end{equation}
Again, $R_{ij}$ is dimensionless and does not depend on the size of
the lens: $R_{ij}\sim 0$ means the images are roughly at the same
distance from the lens galaxy, while $R_{ij}\sim 1$ indicates very
asymmetric configurations that one image lies very close to the lens
center and the other image is far apart from the lens. The other
parameter we use is the opening angle of images:
\begin{equation}
 \theta_{ij}\equiv \cos^{-1}\left(\frac{{\mathbf x}_i
\cdot{\mathbf x}_i} {r_ir_j}\right).
\label{eq:def_tij}
\end{equation}
In this definition, if the images are directly opposite each other,
$\theta_{ij}\sim 180^\circ$. On the other hand, close image pairs such
as merging images near cusp and fold catastrophe have $\theta_{ij}\sim
0^\circ$. Note that both $R_{ij}$ and $\theta_{ij}$ are observables in
the sense that they can be derived without ambiguity for each observed
lens system as long as the lens galaxy is identified: We do not have to
perform mass modeling to compute these quantity from observations.
In summary, our task of this paper is to explore model dependences of
reduced time delay $\Xi$ as a function of image configurations
parameterized by $R_{ij}$ and $\theta_{ij}$. 

\section{Analytic Examination}\label{sec:ana}

In this section, we analytically examine the behavior of the reduced
time delay $\Xi$ (eq. [\ref{eq:xi_def}]) for various lens potentials,
before showing the distribution of $\Xi$ for realistic complicated mass
distributions. For this purpose, it is convenient to study in terms of
multipole expansion: We consider  the lens potential of the following
from 
\begin{equation}
 \phi({\mathbf x})=\sum\frac{c_n}{\beta}R_{\rm Ein}^{2-\beta}r^\beta\cos 
n(\theta-\theta_n),
\end{equation}
where $c_n$ is the dimensionless amplitude and $\theta_n$ is the position
angle. The coefficients are chosen such that $R_{\rm Ein}$ becomes the
Einstein radius of the system if the amplitude of the monopole term is
$c_0=1$. Note that an external shear
\citep[e.g.,][]{kochanek91,keeton97a} can be described by this
expression as $\beta=n=2$ and $c_n=-\gamma$.
For this potential the reduced time delay is given by
\citep{witt00}: 
\begin{equation}
 \Xi=\left|1+\sum 2(1-\beta)\frac{\phi({\mathbf x}_j)-\phi({\mathbf
 x}_i)}{r_j^2-r_i^2}\right|.
\end{equation}
This can be rewritten as
\begin{equation}
 \Xi=\left|1+\sum c_n\left(\frac{2R_{\rm
 Ein}}{r_j+r_i}\right)^{2-\beta}X(R_{ij},\theta_{ij})
\cos(n\theta_n-\delta)\right|,
\label{eq:def_xi}
\end{equation}
where $\delta$ and $X(R_{ij},\theta_{ij})$ are defined by
\begin{equation}
 \tan\delta\equiv \frac{r_j^\beta\sin n\theta_j-r_i^\beta\sin n\theta_i}
{r_j^\beta\cos n\theta_j-r_i^\beta\cos n\theta_i},
\end{equation}
\begin{eqnarray}
X(R_{ij},\theta_{ij})&\equiv&\frac{1-\beta}{2\beta}\frac{1}{R_{ij}}
\left[(1+R_{ij})^{2\beta}+(1-R_{ij})^{2\beta}\right.\nonumber\\
 &&\left.-2(1-R_{ij}^2)^\beta\cos n\theta_{ij}\right]^{1/2}.
\label{eq:def_x}
\end{eqnarray}
Here we assumed $r_j>r_i$ without loss of generality. Note that $R_{ij}$
and $\theta_{ij}$ were defined in equations (\ref{eq:def_rij}) and
(\ref{eq:def_tij}), respectively.  

The above expression of $\Xi$ (eq. [\ref{eq:def_xi}]) has several
important implications. First, $\theta_n$ comes only in the last cosine
term, thus assuming $n\neq 0$ and $c_n$ is small $\Xi>1$ and $\Xi<1$
occurs equally if $\theta_n$ is uncorrelated with the image
configuration (as we will see later, however, this is not necessarily the
case). This indicates that we can reduce the effect of multipole terms
by averaging many lens systems, suggesting the usefulness of our
statistical approach. Second, since $2R_{\rm E}/(r_j+r_i)\sim 1$ in most
cases, the dependence of the potential ($\beta$) and image
configurations ($R_{ij}$ and $\theta_{ij}$) on the deviation from
$\Xi=1$ is encapsulated in $X(R_{ij},\theta_{ij})$, aside from the
overall amplitude $c_n$. In what follows, we use $X(R_{ij},\theta_{ij})$
to study the behavior of $\Xi$. 

First, we consider quite symmetric cases ($R_{ij}\rightarrow 0$) for
which images are nearly the same distance from the lens center. 
Depending on the opening angles, limiting behaviors are given by
\begin{equation}
X(R_{ij}\rightarrow 0,\theta_{ij})\approx \left\{\begin{array}{ll}
{\displaystyle 1-\beta} & (\cos n\theta_{ij}=1)\\
{\displaystyle \frac{1-\beta}{\sqrt{2}\beta R_{ij}}} & (\cos n\theta_{ij}=0)\\
{\displaystyle \frac{1-\beta}{\beta R_{ij}}} & (\cos n\theta_{ij}=-1).
\end{array}\right.
\end{equation}
Therefore, $X(R_{ij},\theta_{ij})$ diverges unless $\cos n\theta_{ij}=1$. 
More rigorously, we should take the limit of  $R_{ij}\rightarrow 0$
and $\theta_{ij}\rightarrow 0$:
\begin{equation}
X(R_{ij}\rightarrow 0,\theta_{ij}\rightarrow 0)\approx
(1-\beta)\left[1+
\left(\frac{n\theta_{ij}}{2\beta R_{ij}}\right)^2\right]^{1/2},
\end{equation}
indicating that the divergence at the symmetric limit can be avoided
if $\theta_{ij}\ll R_{ij}$, or more appropriately $1-\cos
n\theta_{ij}\ll R_{ij}^2$. Since close image pairs are always near the
circumference of the critical curve that has a nearly circular shape
centered on the lens galaxy, such pairs in general have
$R_{ij}\ll\theta_{ij}$, indicating the divergence cannot be avoided for
two close images. On the other hand, the divergence may not occur for
opposite images ($\theta_{ij}\sim 180^\circ$), but only if $n$ is even
number.  

Inversely, if images are very asymmetric ($R_{ij}\rightarrow 1$), 
$X(R_{ij},\theta_{ij})$ reduces to the following simple form
\begin{equation}
X(R_{ij}\rightarrow 1,\theta_{ij})\approx \frac{2^{\beta-1}(1-\beta)}{\beta}.
\end{equation}
This does not depends on $\theta_{ij}$, thus the opening angle is no
longer important in this situation. However it shows stronger
dependence on the radial slope $\beta$.

\begin{figure}
\epsscale{.95}
\plotone{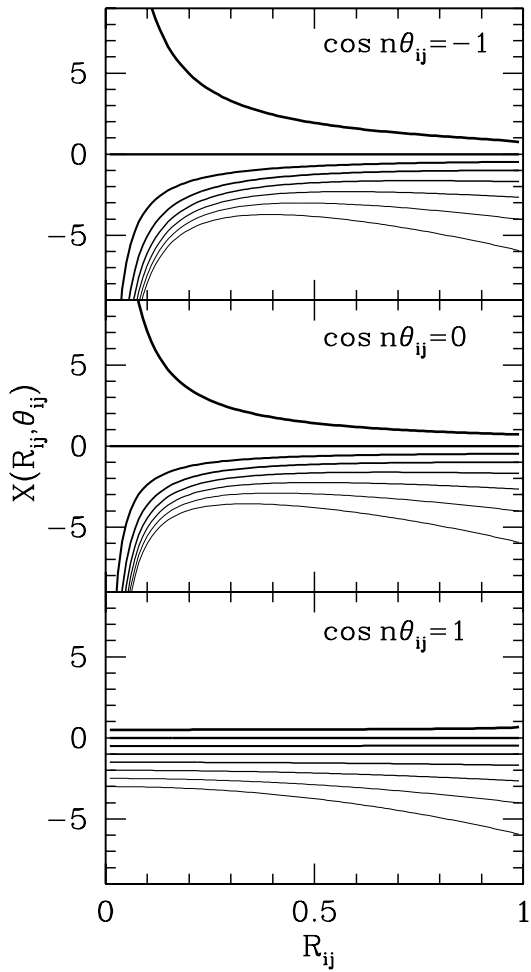}
\caption{The behaviors of $X(R_{ij},\theta_{ij})$
  (eq. [\ref{eq:def_x}]) as a function of the asymmetry parameter
  $R_{ij}$. From thick to thin lines, the radial slope $\beta$ is
  changed from $0.5$ to $4.0$. From top to bottom panels, 
 the opening angle $\theta_{ij}$ is fixed to $\cos n\theta_{ij}=-1$, 
$0$, and $1$, respectively.
\label{fig:x_ana_r}}
\end{figure}

\begin{figure}
\epsscale{.95}
\plotone{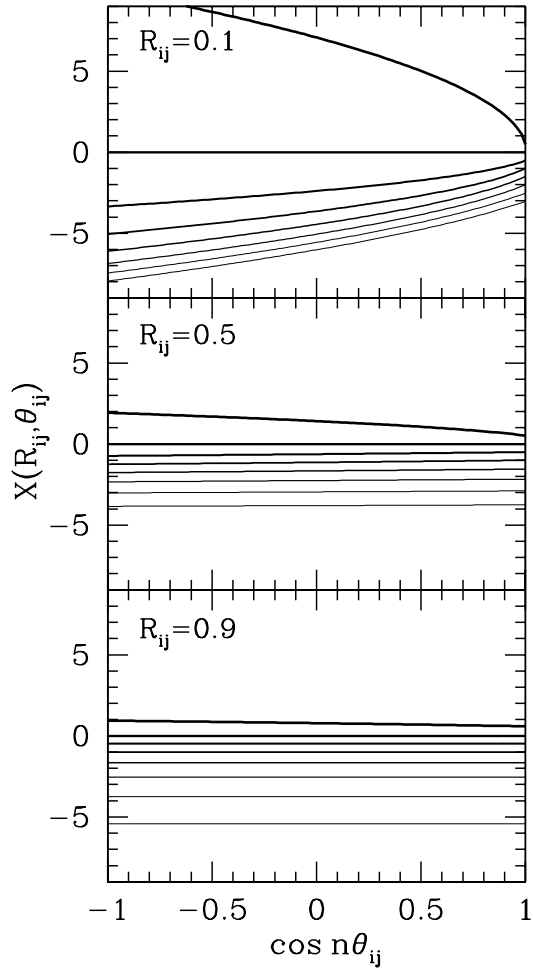}
\caption{Same as Figure \ref{fig:x_ana_r}, but $X(R_{ij},\theta_{ij})$
  is plotted as a function of the opening angle $\theta_{ij}$ while
  fixing $R_{ij}=0.1$ ({\it top panel}), $0.5$ ({\it middle panel}), 
  and $0.9$ ({\it bottom panel}).
\label{fig:x_ana_t}}
\end{figure}

Finally we plot $X(R_{ij},\theta_{ij})$ for various parameter values
in Figures \ref{fig:x_ana_r} and \ref{fig:x_ana_t}. In Figure
\ref{fig:x_ana_r},  $X(R_{ij},\theta_{ij})$ is plotted as a function
of the asymmetry $R_{ij}$. It grows quite rapidly as $R_{ij}$ approaches
to zero for $\cos n\theta_{ij}\neq 1$. It also shows stronger
sensitivity on the slope $\beta$ at larger $R_{ij}$. As is clear in
Figure \ref{fig:x_ana_t},  the opening angle $\theta_{ij}$ becomes
more important for more symmetric lenses. All these behaviors are
consistent with analytical 
arguments presented above. 

\section{Time Delays in Realistic Lens Models}\label{sec:pot}

In this section, we consider more realistic situations to in order to
study expected spread of the reduced time delay $\Xi$ as a function of
image configurations. Specifically we adopt theoretically and
observationally determined distributions of lens potentials such as
ellipticities, external shear, substructures, and multipole components
to make predictions on realistic probability distributions of $\Xi$. 
The methodology is similar to that in \citet{keeton03a,keeton05} who
studied cusp and fold relations to identify lenses with small-scale
structure. 

\begin{figure*}
\epsscale{.95}
\plotone{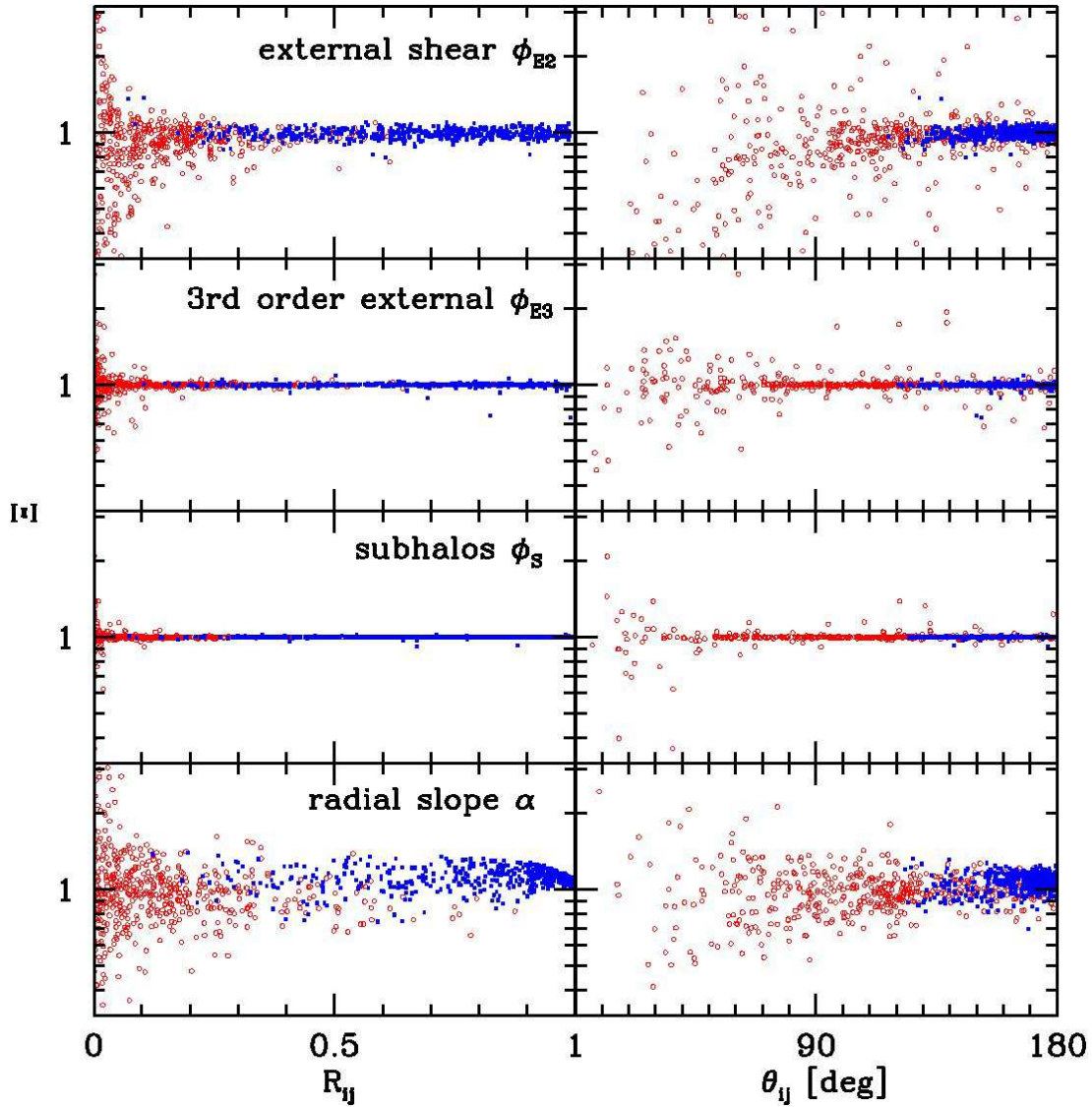}
\caption{Dependences of the reduced time delay $\Xi$ on several lens
 potentials as a function of asymmetry $R_{ij}$ ({\it left panels}) or
 opening angle $\theta_{ij}$ ({\it right panels}). In each panel,
 reduced time delays of 500 double lenses ({\it filled squares}) and
 500 quadruple lenses ({\it open circles}) obtained by the Monte-Carlo
 simulations are plotted. Magnification bias is not considered
 at this stage. From top to bottom panels, we consider the external
 shear $\phi_{\rm E2}$ (eq. [\ref{eq:pot_e2}]), third order external
 perturbation $\phi_{\rm E3}$ (eq. [\ref{eq:pot_e3}]), subhalos
 $\phi_{\rm S}$ (eq. [\ref{eq:pot_s}]), and non-isothermality
 $\alpha\neq 1$ in the primary lens potential  $\phi_{\rm G}$
 (eq. [\ref{eq:pot_g}]). Besides the bottom panels, we adopt an
 isothermal elliptical lens as primary lens galaxies
 (eq. [\ref{eq:pot_g}], $\alpha=1$) and ignore the other perturbations
 (e.g., in the second row we only consider $\phi_{\rm E3}$ and
 ignore the external shear and subhalos), thus the effect of each
 potential can be measured by the deviation from $\Xi=1$. In all
 simulations higher order multipoles (eq. [\ref{eq:pot_m}]) are included.
\label{fig:xi_model}}
\end{figure*}

\subsection{Input Models}\label{sec:pot_input}
As primary  lens galaxies we only consider elliptical galaxies because
most ($>80$\%) of quasar lenses are caused by massive elliptical
galaxies \citep[e.g.,][]{turner84,kochanek06a,moeller07}. We model the
lens galaxy as a power-law elliptical mass distribution. The surface mass
distribution is given by   
\begin{equation}
 \kappa_{\rm G}({\mathbf x})=\frac{\alpha}{2}\left[\frac{R_{\rm Ein}}
{r\sqrt{1-\epsilon\cos2(\theta-\theta_e)}}\right]^{2-\alpha},
\end{equation}
where $R_{\rm Ein}$ is the Einstein ring radius and $\theta_e$ is the
position angle of ellipse. The case $\alpha=1$ corresponds to the
standard singular isothermal mass distribution. The ellipticity $e$,
which is defined by $e=1-q$ with $q$ being the axis ratio of
the ellipse, is related to $\epsilon$ by 
\begin{equation}
 \epsilon=\frac{1-(1-e)^2}{1+(1-e)^2}.
\end{equation}
The corresponding lens potential can be described by
\begin{equation}
 \phi_{\rm G}({\mathbf x})=\frac{1}{\alpha}R_{\rm
 Ein}^{2-\alpha}r^\alpha G(\theta),
\label{eq:pot_g}
\end{equation}
with $G(\theta)$ being the complex function of $\theta$, but notice
that $G(\theta)=1$ if $e=0$. 

Many previous work has shown that lens galaxies indeed have nearly
isothermal mass distribution. \citet{rusin05} obtained $\alpha=0.94\pm
0.17$ by combining the Einstein radii of many lensed quasars with the
fundamental plane relation of elliptical galaxies. The slope of the
galaxy density profile $\alpha=1.09\pm0.01$ constrained from the faint
third image of PMN J1632$-$0033 is consistent with the nearly
isothermal density profile \citep{winn04}. Detailed mass modeling of 
B1933+503 also indicates nearly isothermal profile of the lens galaxy
\citep{cohn01}. 
By using measured velocity dispersions of lens galaxies of several 
lensed quasars, \citet[][see also \citealt{hamana07}]{treu04} derived
$\alpha=1.25\pm0.2$. \citet{koopmans06} put tighter constrains
$\alpha=0.99^{+0.03}_{-0.02}$, but the results are derived from a
sample of much lower redshift lens systems than typical time delay 
quasars. From these results, in this paper we adopt the Gaussian
distribution $\alpha=1\pm0.15$ as a conservative input distribution of
the slope.\footnote{The adopted distribution does not include the mean
value of \citet{treu04} within its 1-$\sigma$ uncertainty. However, we
note that the result of \citet{treu04} were drawn from only five lensed
quasar systems and therefore scatter should be large. Moreover, many
of lens systems used by \citet{treu04} appear to reside in dense
environments, which may explain their large value of $\alpha$. Indeed,
\citet{rusin05} used a larger sample of 22 lenses to derive the slope
that is more consistent with our input distribution. } For
the ellipticity, we use the Gaussian distribution with  median $e=0.3$
and dispersion $0.16$ that is consistent with observed distributions
of isodensity contour shapes of elliptical galaxies
\citep{bender89,saglia93,jorgensen95,rest01,sheth03}.    

To allow more complex mass distribution of the galaxy, we add higher
order multipole terms to the potential: 
\begin{equation}
 \phi_{\rm M}({\mathbf x})=\frac{1}{\alpha}R_{\rm
 Ein}^{2-\alpha}r^\alpha\sum_m(1-m^2)A_m\cos m(\theta-\theta_m).
\label{eq:pot_m}
\end{equation}
The factor $1-m^2$ is inserted such that $A_m$ denotes the standard
parametrization for the deviation of the mass density from an
ellipsoid. We include only $m=3$ and $4$ terms, because $m\geq 5$
perturbations have generally not been reported. For both $A_3$ and
$A_4$, the amplitudes are distributed by the Gaussian with mean zero
and dispersion $0.01$ that is roughly consistent with reported
distributions \citep{bender89,saglia93,rest01}. It is also compatible
with the level of the deviation inferred from individual modeling of
lensed quasar systems \citep{trotter00,kawano04,kochanek04,congdon05,yoo06}. 
We assume that position angles of these multipole perturbations and
that of an ellipsoid are uncorrelated with each other. In addition, we
neglect the correlation of $e$ and $A_4$ \citep{keeton03a} for simplicity. 
Note that these multipole perturbations do not affect $\Xi$ if the
lens galaxy is isothermal, $\alpha=1$, but can change quantitative
results when combined with non-isothermal profiles and/or other
small-scale structures. Since the effect of $A_4$ itself is small, we
expect the effect of the correlation between $e$ and $A_4$, which we
have neglected here, is also small. 

We also include external perturbations that are known to be important
for individual mass modeling \citep[e.g.,][]{keeton97a}. The potential
of lowest order external shear is given by
\begin{equation}
 \phi_{\rm E2}({\mathbf
  x})=-\frac{\gamma}{2}r^2\cos2(\theta-\theta_\gamma).
\label{eq:pot_e2}
\end{equation}
We adopt a log-normal distribution with median $\gamma=0.05$ and
dispersion 0.2 dex for the distribution of shear amplitude. It is
consistent with expected shear distribution from $N$-body simulations 
\citep{holder03,dalal04}. In addition to external shear, we consider third
order external perturbation \citep{kochanek91,bernstein99}
\begin{equation}
 \phi_{\rm E3}({\mathbf x})=\frac{\sigma}{4}R_{\rm Ein}^{-1}
r^3\left[\cos(\theta-\theta_\sigma)-\cos3(\theta-\theta_\sigma)\right].
\label{eq:pot_e3}
\end{equation}
Here we assumed the external perturber is a singular isothermal
object to relate the amplitudes and position angles of $n=1$ and $n=3$
terms. In general we have $\sigma\approx \gamma^2$
\citep{bernstein99}, thus for the amplitude we adopt a log-normal
distribution with median $\sigma=\gamma^2$ and dispersion 0.2 dex.
The adopted amplitude is roughly consistent with values obtained from
mass modeling of individual lens systems \citep[e.g.,][]{kawano04}.
We allow small mis-alignment of the position angle $\theta_\sigma$,
which could happen when the external perturber is not spherical, by
adopting the Gaussian distribution around $\theta_\gamma$ with
dispersion $10^\circ$.

\begin{figure*}
\epsscale{.95}
\plotone{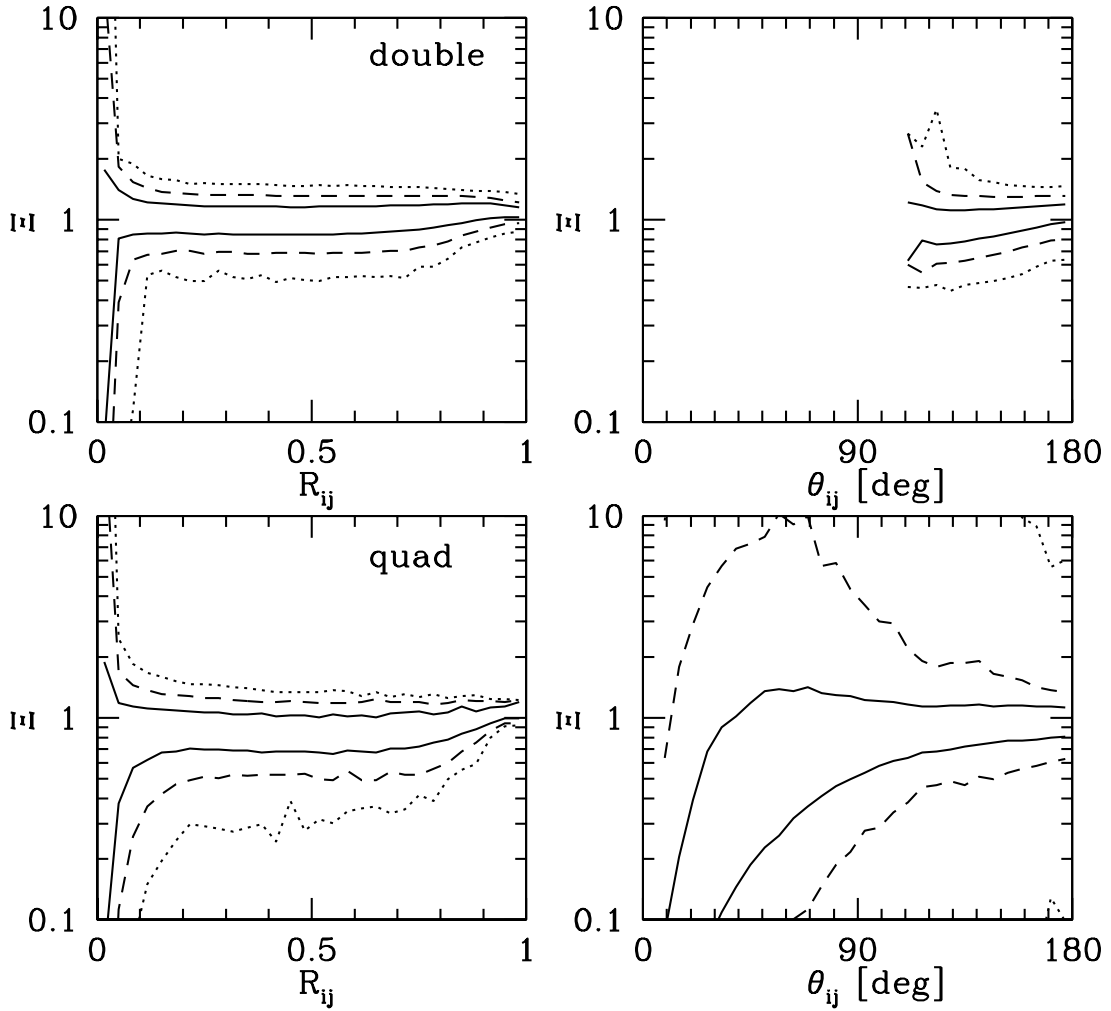}
\caption{Contours of constant conditional probability $p(\Xi|R_{ij})$
  ({\it left panels}) and $p(\Xi|\theta_{ij})$ ({\it right panels})
  computed from  Monte-Carlo simulations of realistic lens
  potentials. The contours are drawn at the 68\% ({\it solid lines}),
  95\% ({\it dashed lines}), and 99.7\% ({\it dotted lines})
  confidence levels. Upper panels show the conditional probability for
  double lenses, whereas lower panels are for quadruple lenses. For
  double lenses we show $p(\Xi|\theta_{ij})$ only at
  $\theta_{ij}\gtrsim 110^\circ$ because we have too few image pairs
  with $\theta_{ij}\lesssim 110^\circ$ to construct the probability
  distribution.  
\label{fig:dist}}
\end{figure*}

Finally we consider substructures in the lens galaxy. A significant
fraction of substructures (subhalos) in galaxies have been predicted
as a natural consequence of cold dark matter cosmology
\citep{moore99,klypin99}. Anomalous flux ratios observed in many
gravitational lens systems indicate that substructures are indeed
present \citep{metcalf01,chiba02,dalal02,bradac02,keeton05,
kochanek04,metcalf04,chiba05}.
Note that small perturbations may also come from small halos along the
line-of-sight \citep{keeton03b,chen03,oguri05a,metcalf05}. 
Although time delays are thought to be less sensitive to small-scale
structure than flux ratios that are determined by the second
derivative of the time delay surface, they might be affected to some
extent particularly when two images are close to each other.
We model each subhalo by pseudo-Jaffe (truncated singular isothermal)
profile. The lens potential of this profile is
\begin{eqnarray}
 \phi_{{\rm PJ},k}({\mathbf
 x})&=&b_k\left[r-\sqrt{r^2+a^2_k}\right.\nonumber\\
 &&\left.-\frac{a_k}{2}\ln\left|
\frac{\sqrt{r^2+a^2_k}-a_k}{\sqrt{r^2+a^2_k}+a_k}\right|+a_k\ln
 r\right],
\end{eqnarray}
where $a_k$ is a truncation radius and $b_k$ is a mass
normalization that coincides the Einstein radius for sufficiently large
$a_k$. We adopt $a_k=\sqrt{b_kR_{\rm E}}$ assuming the truncation
radius of the tidal radius of the subhalo \citep[see, e.g.,][]{metcalf01}.
For the velocity distribution  we assume $N(>v)=(10v/V)^{-2.7}$ inside
three times the Einstein radius of the lens galaxy, where $v$ and $V$
are velocity dispersions of the subhalo and halo, respectively. The
velocities can be converted to $b_k$ through $b_k/R_{\rm E} \propto
(v/V)^2$. We distribute subhalos randomly with an uniform spatial
density in the projected two-dimensional plane in order to take
account of the suggested anti-bias of the subhalo spatial distribution
\citep{delucia04,mao04,oguri04b}. The resulting mass fraction of
subhalos at around the Einstein radius is $\sim 0.5$\%, being consistent
with the expectation from $N$-body simulations and analytic calculation
\citep[e.g.,][]{mao04,oguri05a}. The effect of subhalos is described by
the sum of the lens potential of each subhalo: 
\begin{equation}
 \phi_{\rm S}({\mathbf x})=\sum_k\phi_{{\rm PJ},k}({\mathbf x}-{\mathbf
 x}_{{\rm sub},k})-\frac{1}{2}r^2\bar{\kappa}_{\rm sub}.
\label{eq:pot_s}
\end{equation}
Here we subtracted the convergence averaged over all subhalos,
$\kappa_{\rm sub}$, to conserve the total mass and radial profile of
the lens galaxy. We include only 100 most massive subhalos mainly for
computational reason, but small subhalos are expected to have
only very small effect on time delays.

\begin{figure*}
\epsscale{.95}
\plotone{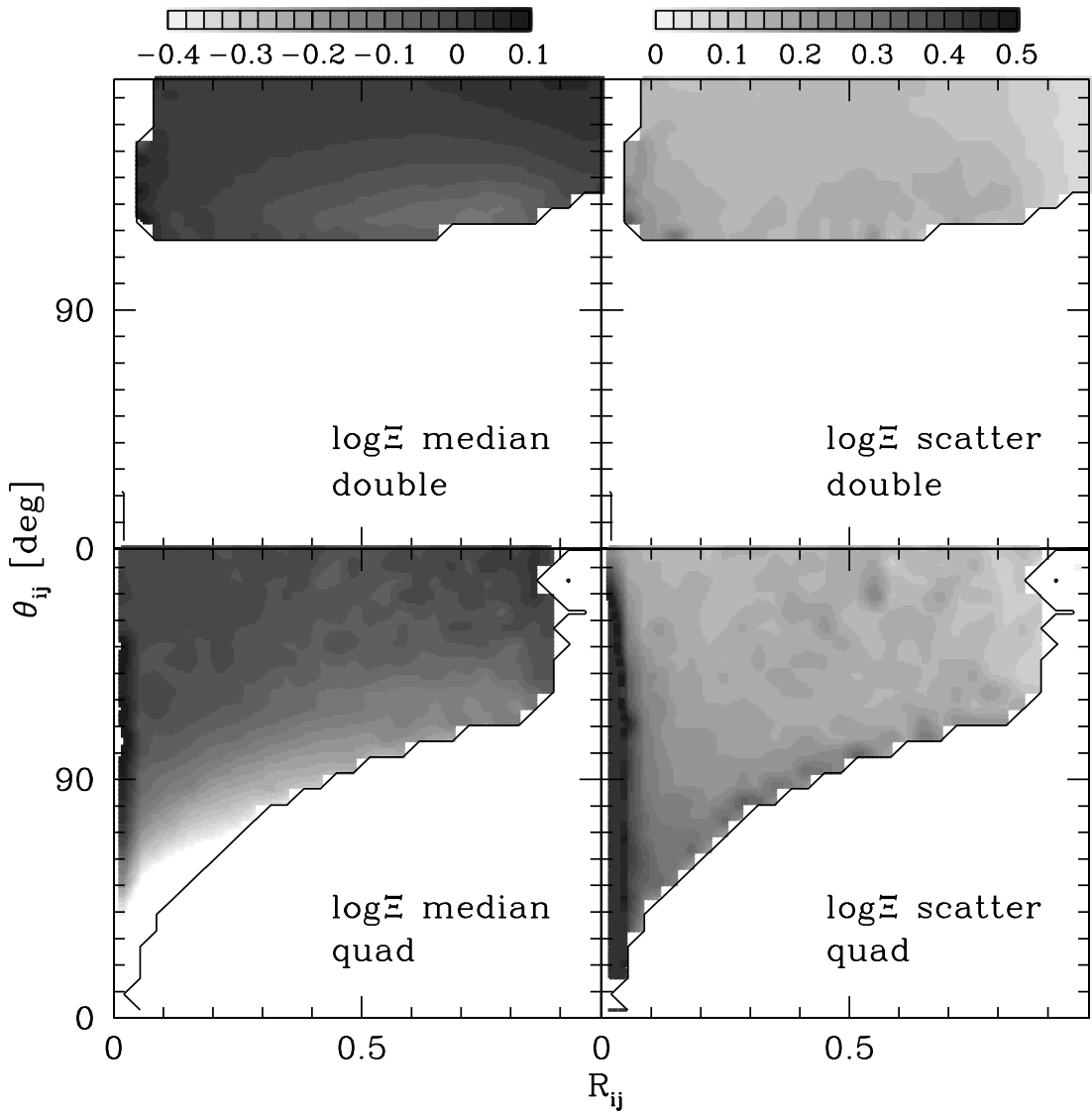}
\caption{Contour plot of median ({\it left panels}) and scatter ({\it
  right panels}) of the  conditional probability
  $p(\Xi|R_{ij},\theta_{ij})$ in the $R_{ij}$-$\theta_{ij}$
  plane. Here the scatter is defined by the 68\% confidence interval
  width in units of $\log\Xi$. The probability distribution for double
  ({\it upper panels}) and quadruple ({\it lower panels}) lenses are
  shown separately. Thick solid lines indicate the limit beyond which
  the number of image pairs in Monte-Carlo simulations is too small to
  construct the conditional probability. 
\label{fig:cont}}
\end{figure*}

\subsection{Simulation Method}\label{sec:pot_sim}

Using the model described above, we perform a large Monte-Carlo
simulation containing $>10^6$ lensed image pairs.
The simulation is done by first generating a lens potential according
to the distributions summarized in \S \ref{sec:pot_input}. All lengths
are scaled by the Einstein ring radius $R_{\rm E}$: Since our results
do not depend on adopted length scales, we fix $R_{\rm E}$ to unity in
the simulations. We generate 10000 different lens potentials in total. For
each lens potential, we place random sources with an uniform density
of $\sim 100R_{\rm   E}^{-2}$ in the source plane. We use a public
software {\it   lensmodel}  \citep{keeton01} to solve the lens
equation and compute time delays between multiple images. The uniform
sampling in the source plane indicates that each lens potential is
automatically weighted by the lensing cross section
\citep[see][]{keeton03a}. To account for magnification bias as well,
for each source we compute the total magnification factor $\mu_{\rm
  tot}$, and when constructing probability distributions of reduced
time delays (see below) we include a weight of $\mu_{\rm tot}^{q-1}$,
where $q$ is a power-law slope of the luminosity function of source
quasars. We adopt $q=2.1$ that is relevant for lenses identified by
the CLASS \citep{myers03}.  

From the ensemble of image pairs, we compute conditional
probability distribution functions of the reduced time delay $\Xi$ for
given values of the asymmetry $R_{ij}$ and/or opening angle
$\theta_{ij}$, i.e., $p(\Xi|R_{ij})$, $p(\Xi|\theta_{ij})$, and
$p(\Xi|R_{ij}, \theta_{ij})$. The probability distributions are
computed separately for double and quadruple lenses to see how different
distributions they exhibit. In what follow we ignore central faint
images that are unobserved in most cases. 

\begin{figure*}
\epsscale{.95}
\plotone{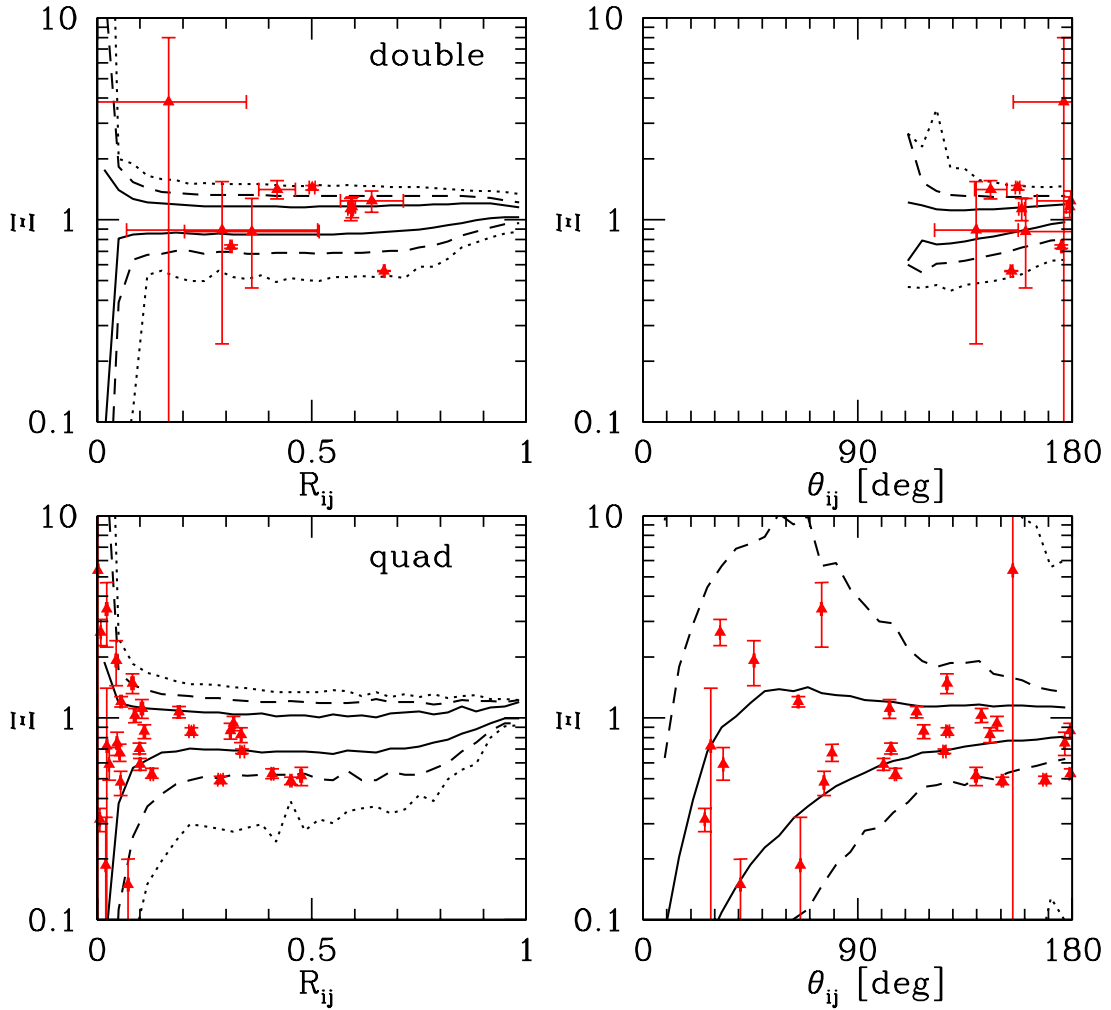}
\caption{Same as Figure \ref{fig:dist}, but observed values of $\Xi$
 with error-bars (see Table \ref{tab:xi}) are plotted on the contours
 by filled triangles with error-bars. The Hubble constant of $h=0.73$
 is assumed when computing $\Xi$ from observed time delays.
\label{fig:dist_obs}}
\end{figure*}

\subsection{Contribution of Each Potential}\label{sec:pot_pot}
Before presenting results that include all potentials, it is useful to
see how each potential affects the distribution of $\Xi$. To see this we
adopt isothermal elliptical lens potential plus multipole terms
($\phi_{\rm G}+\phi_{\rm M}$, $\alpha=1$), and add each lens potential:
Since the isothermal galaxy always yields $\Xi=1$ (see \S \ref{sec:model})
the effect of each potential term can be estimated by the deviation from
$\Xi=1$. We also study the effect of non-isothermality ($\alpha\neq1$).
The results are summarized in Figure \ref{fig:xi_model}. First,
the effect of external shear appears as a scatter around $\Xi=1$,
which is consistent with discussion above (\S \ref{sec:ana}) as well
as that of \citet{witt00}. Both the diverging scatter at
$R_{ij}\rightarrow 0$ and the modest increase of the scatter at
smaller $\theta_{ij}$ are expected from our analytic examination. 
The scatter introduced by the third order external perturbation and
subhalos is smaller than that of external shear, but it is still
noticeable particularly for symmetric configurations. This means that
even if we can estimate external shear and its orientation accurately
for individual lens system by detailed mass modeling, these third order
perturbations or subhalos can shift time delays, although these small
perturbations mainly affect image pairs with small opening angles and
only increase the scatter rather than shift the mean.
We comment that the use of image pairs with small opening
angles is also limited by the larger uncertainties of their time
delay measurements. As expected, non-isothermality has a large impact
on $\Xi$, but the size of its scatter is less dependent on image
configurations compared with other  potentials.

It is worth noting that there are systematic deviations from $\Xi=1$
for a few specific image configurations. For instance, at small
$\theta_{ij}$  external shear preferentially produces image pairs with
time delays smaller than the isothermal case. In addition, very
asymmetric image pairs ($R_{ij}\sim 1$) tend to have larger time
delays when we consider non-isothermal lens galaxies. Such systematic
shifts were not predicted in our analytic arguments in
\S \ref{sec:ana}, thus invite careful consideration. We find that 
the selection effect can explain these systematic shifts. For very
asymmetric pairs, such configurations are possible only when the lens
galaxy has a profile steeper than isothermal for which inner critical
curve is degenerated at the center. If the density profile is shallower
than isothermal, any images (expect central faint images that we
ignore in this paper) are outside the inner critical curve.
This means that very asymmetric configuration, in which one image lies
very close to the center of the lens potential, never occurs for the
profile shallower than isothermal. Since shallower profiles correspond
to smaller $\Xi$, this increases the mean $\Xi$ at $R_{ij}\sim 1$.
The effect of external shear is more complicated, but can be
explained as follows. Consider a close pair of two images. If we add
external shear whose position angle direction is perpendicular to the
segment that connects the two images, it separates two images. On the
other hand, external shear that is parallel to the segment makes the two
images closer. Therefore, combined with the steep dependence of the
frequency of close image pairs on the opening angle, this results in
the situation that at fixed small opening angle the direction of
external shear is more likely to be parallel to the segment than
perpendicular. The discussion in \S  \ref{sec:ana} indicates that the
parallel shear yields smaller time delays $\Xi<1$ for fixed image
positions, thus we can conclude that close image pairs have
statistically smaller time delays than those expected without external
shear.    

\subsection{Conditional Probability Distributions}\label{sec:pot_xi}

We are now ready to compute conditional probability distributions of
the reduced time delay $\Xi$ with all lens potentials presented in \S
\ref{sec:pot_input}. Figure \ref{fig:dist} shows contours of constant
conditional probability projected to one-dimensional surface,
$p(\Xi|R_{ij})$ and $p(\Xi|\theta_{ij})$. The behaviors of contours
can well be understood from discussions in \S \ref{sec:ana} and
\S \ref{sec:pot_pot}. In particular, systematically small $\Xi$ at
small opening angles and large $\Xi$ for very asymmetric image pairs,
which we discussed in \S \ref{sec:pot_pot}, are clearly seen. It
appears quadruple lenses have larger scatter on average: This is
because strong perturbations (external shear, third order perturbation
and subhalos), which cause the scatter around $\Xi=1$, also enhance
the lensing cross section for quadruple images. 

To study the conditional probability distribution as a function of
both $R_{ij}$ and $\theta_{ij}$, $p(\Xi|R_{ij},\theta_{ij})$, 
in Figure \ref{fig:cont} we plot contours of median and 68\% confidence
interval of the probability distribution in the $R_{ij}$-$\theta_{ij}$
plane. The features we have discussed, such as the divergence of
scatter at $R_{ij}\rightarrow 0$, smaller median time delays for small
opening angle pairs, larger median time delays for asymmetric pairs,
can be seen in this Figure as well. In addition, the scatter presented
here is useful to check which lens systems are less dependent on lens
potentials: Our result indicates that asymmetric image pairs that are
collinear with their lens galaxies ($\theta_{ij}\sim 180^\circ$) have
the least scatter and therefore are more suitable for the determination
of the Hubble constant. It is interesting to note that in the
statistical sense double lenses are more valuable than quadruple lenses
because it shows a smaller sensitivity to various lens potentials. In
addition, double lenses have advantages of easier measurements of
time delays in observations and smaller fractional errors (because
double lenses have longer time delays than quads). This is in contrast
to individual mass modeling in which quadruple lenses are more useful
because of much more observational constraints on mass models. 

\begin{deluxetable*}{lllllllccl}
\tablewidth{0pt}
\tabletypesize{\footnotesize}
\tablecaption{Summary of Observed Quasar Time Delays}
\tablewidth{0pt}
\tablehead{
 \colhead{Lens Name} & \colhead{$N_{\rm img}$} & \colhead{$z_s$} &
 \colhead{$z_l$} & \colhead{Images} & 
 \colhead{$R_{ij}$} & \colhead{$\theta_{ij}$ [deg]} & 
 \colhead{$\Delta t$ [days]} & \colhead{$\Xi(h=0.73)$\tablenotemark{a}} & 
 \colhead{References}}
\startdata
     B0218+357 & 2 & 0.944 & 0.685 & AB &
$0.167\pm0.181$ & $176.5\pm 21.3$ &
$ 10.5\pm 0.2$ & $3.835\pm4.155$ &
    1, 2, 3, 4, 5\\
   HE0435$-$1223 & 4 & 1.689 & 0.455 & AD &
$0.099\pm0.002$ & $103.9\pm  0.2$ &
$ 14.4^{+0.9}_{-0.8}$ & $0.704^{+0.045}_{-0.041}$ &
   6, 7, 8\\
   &&& & AB &
$0.053\pm0.001$ & $ 79.3\pm  0.2$ &
$  8.0^{+0.8}_{-0.7}$ & $0.671^{+0.069}_{-0.060}$ &
    \\
   &&& & AC &
$0.002\pm0.002$ & $155.1\pm  0.2$ &
$  2.1^{+0.7}_{-0.8}$ & $5.360^{+6.070}_{-6.150}$ &
    \\
   &&& & BD &
$0.046\pm0.002$ & $176.8\pm  0.3$ &
$  6.4\pm 0.8$\tablenotemark{c} & $0.750\pm0.100$ &
    \\
   &&& & CD &
$0.100\pm0.002$ & $100.9\pm  0.3$ &
$ 12.3\pm 0.8$\tablenotemark{c} & $0.590\pm0.040$ &
    \\
   &&& & BC &
$0.054\pm0.001$ & $ 75.9\pm  0.3$ &
$  5.9\pm 0.8$\tablenotemark{c} & $0.479\pm0.066$ &
    \\
  RXJ0911+0551 & 4 & 2.800 & 0.769 & A1B &
$0.452\pm0.003$ & $150.8\pm  0.4$ &
$143.0\pm 6.0$ & $0.488\pm0.021$ &
   9, 10, 11, 12 \\
  &&& & A2B &
$0.408\pm0.003$ & $179.3\pm  0.5$ &
$149.0\pm 8.0$ & $0.530\pm0.029$ &
   \\
  &&& & A3B &
$0.476\pm0.003$ & $139.6\pm  0.5$ &
$154.0\pm16.0$ & $0.516\pm0.054$ &
   \\
  SBS0909+532 & 2 & 1.377 & 0.830 & AB &
$0.291\pm0.223$ & $139.7\pm 17.6$ &
$ 45.0^{+5.5}_{-0.5}$ & $0.891^{+0.657}_{-0.648}$ &
    4, 13, 14, 15\\
  FBQ0951+2635 & 2 & 1.246 & 0.260 & AB &
$0.591\pm0.007$ & $158.8\pm  1.2$ &
$ 16.0\pm 2.0$ & $1.134\pm0.142$ &
    12, 16, 17, 18\\
     Q0957+561 & 2 & 1.413 & 0.36 & AB &
$0.669\pm0.002$ & $154.5\pm  0.5$ &
$417.0\pm 1.5$ & $0.558\pm0.002$ &
    19, 20, 21, 22\\
SDSS J1004+4112 & 5 & 1.734 & 0.68 & AB\tablenotemark{b} &
$0.006\pm0.001$ & $ 25.9\pm  0.1$ &
$ 38.4\pm 1.0$ & $0.315\pm0.041$ &
    23, 24, 25, 26\\
   HE1104$-$1805 & 2 & 2.319 & 0.729 & AB &
$0.312\pm0.002$ & $175.5\pm  0.3$ &
$152.2^{+2.8}_{-3.0}$ & $0.738^{+0.015}_{-0.016}$ &
   4, 27, 28, 29, 30\\
    PG1115+080 & 4 & 1.735 & 0.310 & A1B &
$0.105\pm0.003$ & $103.4\pm  0.4$ &
$ 11.7\pm 1.2$ & $1.115\pm0.120$ &
   31, 32, 33, 34, 35, 36 \\
    &&& & A2B &
$0.082\pm0.004$ & $127.5\pm  0.4$ &
$ 11.7\pm 1.2$ & $1.487\pm0.166$ &
    \\
    &&& & BC &
$0.191\pm0.003$ & $114.7\pm  0.3$ &
$ 25.0\pm 1.6$ & $1.069\pm0.070$ &
    \\
    &&& & A1C &
$0.088\pm0.003$ & $141.9\pm  0.3$ &
$ 13.3\pm 1.0$ & $1.032\pm0.083$ &
    \\
    &&& & A2C &
$0.110\pm0.003$ & $117.8\pm  0.3$ &
$ 13.3\pm 1.0$ & $0.857\pm0.067$ &
    \\
    &&&  & A1A2 &
$0.022\pm0.003$ & $ 24.1\pm  0.3$ &
$  0.149\pm 0.006$ & $0.057\pm0.007$ &
    \\
  RXJ1131$-$1231 & 4 & 0.658 & 0.295 & AB &
$0.008\pm0.001$ & $ 32.3\pm  0.1$ &
$ 12.0^{+1.5}_{-1.3}$ & $2.660^{+0.411}_{-0.376}$ &
   37, 38  \\
  &&& & AC &
$0.028\pm0.000$ & $ 33.7\pm  0.0$ &
$  9.6^{+2.0}_{-1.6}$ & $0.589^{+0.123}_{-0.098}$ &
    \\
  &&& & BC &
$0.020\pm0.001$ & $ 66.0\pm  0.1$ &
$  2.2\pm 1.6$ & $0.187\pm0.136$ &
    \\
  &&& & AD &
$0.311\pm0.002$ & $179.2\pm  0.1$ &
$ 87.0\pm 8.0$ & $0.863\pm0.079$ &
    \\
  &&& & BD &
$0.318\pm0.002$ & $148.5\pm  0.2$ &
$ 99.0\pm 8.0$\tablenotemark{c} & $0.940\pm0.076$ &
    \\
  &&& & CD &
$0.335\pm0.002$ & $145.5\pm  0.1$ &
$ 96.6\pm 8.0$\tablenotemark{c} & $0.825\pm0.068$ &
    \\
     B1422+231 & 4 & 3.620 & 0.337 & AB &
$0.023\pm0.002$ & $ 28.4\pm  0.2$ &
$  1.5\pm 1.4$ & $0.726\pm0.680$ &
   34, 39, 40 \\
   &&& & AC &
$0.022\pm0.003$ & $ 74.9\pm  0.3$ &
$  7.6\pm 2.5$ & $3.461\pm1.216$ &
    \\
   &&& & BC &
$0.045\pm0.003$ & $ 46.5\pm  0.3$ &
$  8.2\pm 2.0$ & $1.924\pm0.482$ &
    \\
   SBS1520+530 & 2 & 1.855 & 0.717 & AB &
$0.501\pm0.007$ & $157.1\pm  0.8$ &
$130.0\pm 3.0$ & $1.444\pm0.037$ &
    41, 42, 43\\
     B1600+434 & 2 & 1.589 & 0.414 & AB &
$0.640\pm0.073$ & $179.4\pm 14.2$ &
$ 51.0\pm 2.0$ & $1.242\pm0.150$ &
    44, 45, 46, 47\\
     B1608+656 & 4 & 1.394 & 0.630 & AB &
$0.127\pm0.004$ & $105.9\pm  0.4$ &
$ 31.5^{+2.0}_{-1.0}$ & $0.525^{+0.037}_{-0.023}$ &
    48, 49, 50, 51, 52\\
     &&& & BC &
$0.056\pm0.002$ & $ 65.1\pm  0.3$ &
$ 36.0\pm 1.5$ & $1.202\pm0.072$ &
    \\
     &&& & BD &
$0.338\pm0.005$ & $126.4\pm  0.6$ &
$ 77.0^{+2.0}_{-1.0}$ & $0.681^{+0.020}_{-0.013}$ &
    \\
     &&& & AC &
$0.072\pm0.002$ & $ 40.8\pm  0.2$ &
$  4.5\pm 1.5$\tablenotemark{c} & $0.150\pm0.050$ &
    \\
     &&& & AD &
$0.220\pm0.006$ & $127.8\pm  0.7$ &
$ 45.5\pm 1.5$\tablenotemark{c} & $0.858\pm0.036$ &
    \\
     &&& & CD &
$0.287\pm0.006$ & $168.5\pm  0.7$ &
$ 41.0\pm 1.5$\tablenotemark{c} & $0.494\pm0.021$ &
    \\
SDSS J1650+4251 & 2 & 1.547 & 0.577 & AB &
$0.420\pm0.043$ & $145.8\pm  6.8$ &
$ 49.5\pm 1.9$ & $1.415\pm0.149$ &
    53, 54\\
   PKS1830$-$211 & 2 & 2.507 & 0.89 & AB &
$0.360\pm0.157$ & $160.5\pm 20.3$ &
$ 26.0^{+4.0}_{-5.0}$ & $0.874^{+0.401}_{-0.414}$ &
   4, 55, 56, 57, 58 \\
   HE2149$-$2745 & 2 & 2.033 & 0.603 & AB &
$0.594\pm0.007$ & $178.9\pm  1.5$ &
$103.0\pm12.0$ & $1.161\pm0.136$ &
   12, 59, 60\\
\enddata
\tablecomments{All measured time delays are listed, except time delays
  between images A1---A3 of RXJ0911+0551 and images A---D of Q2237+030
  \citep{vakulik06} for which error-bars are large and therefore the
 detections are marginal. For Q2237+030, a possible X-ray detection of
 the time delay between image A and B was also reported by
 \citet{dai03}. Errors indicate 1$\sigma$. }  
\tablenotetext{a}{The values of reduced time delays $\Xi$ computed
  from observed time delays and image configurations. For the Hubble
  constant $h=0.73$ is assumed. Errors of $\Xi$
  come from those of $\Delta t$ and $|r_j^2-r_i^2|$. } 
\tablenotetext{b}{We assume the position of the brightest cluster galaxy
 G1 for the center of the lens potential, though mass modeling implies
 the significant offset of the potential center from G1 \citep{oguri04a}.} 
\tablenotetext{c}{The values and errors of time delays were not
  directly given in the literature, thus we inferred them from those
  of other image pairs.} 
\tablerefs{(1) \citealt{patnaik93}; (2) \citealt{browne93}; 
(3) \citealt{biggs99}; (4) \citealt{lehar00}; (5) \citealt{cohen03};
(6) \citealt{wisotzki02}; (7) \citealt{morgan05}; 
(8) \citealt{kochanek06b}; (9) \citealt{bade97}; (10) \citealt{kneib00}; 
(11) \citealt{hjorth02}; (12) CASTLES (http://cfa-www.harvard.edu/castles/);
(13) \citealt{oscoz97}; (14) \citealt{lubin00}; (15) \citealt{ullan06};  
(16) \citealt{schechter98}; (17) \citealt{jakobsson05}; 
(18) \citealt{eigenbrod07}; (19) \citealt{walsh79}; 
(20) \citealt{young81}; (21) \citealt{kundic97}; 
(22) \citealt{barkana99}; (23) \citealt{inada03}; (24) \citealt{oguri04a};
(25) \citealt{inada05}; (26) \citealt{fohlmeister07}; 
(27) \citealt{wisotzki93}; (28) \citealt{lidman00}; (29) \citealt{ofek03}; 
(30) \citealt{poindexter06}; (31) \citealt{weymann80}; 
(32) \citealt{schechter97}; (33) \citealt{barkana97}; 
(34) \citealt{tonry98}; (35) \citealt{impey98}; (36) \citealt{chartas04};
(37) \citealt{sluse03}; (38) \citealt{morgan07}; 
(39) \citealt{patnaik92}; (40) \citealt{patnaik01};
(41) \citealt{chavushyan97}; (42) \citealt{burud02b}; (43) \citealt{faure02};
(44) \citealt{jackson95}; (45)  \citealt{fassnacht98}; 
(46) \citealt{koopmans98}; (47) \citealt{burud00}; (48) \citealt{myers95};
(49) \citealt{fassnacht96}; (50) \citealt{koopmans99}; 
(51) \citealt{fassnacht02}; (52) \citealt{koopmans03}; 
(53) \citealt{morgan03}; (54) \citealt{vuissoz07};
(55) \citealt{subrahmanyan90}; (56) \citealt{wiklind96};
(57) \citealt{lovell98}; (58) \citealt{lidman99};
(59) \citealt{wisotzki96}; (60) \citealt{burud02a}.}

\label{tab:xi}
\end{deluxetable*}

\begin{figure*}
\epsscale{.99}
\plotone{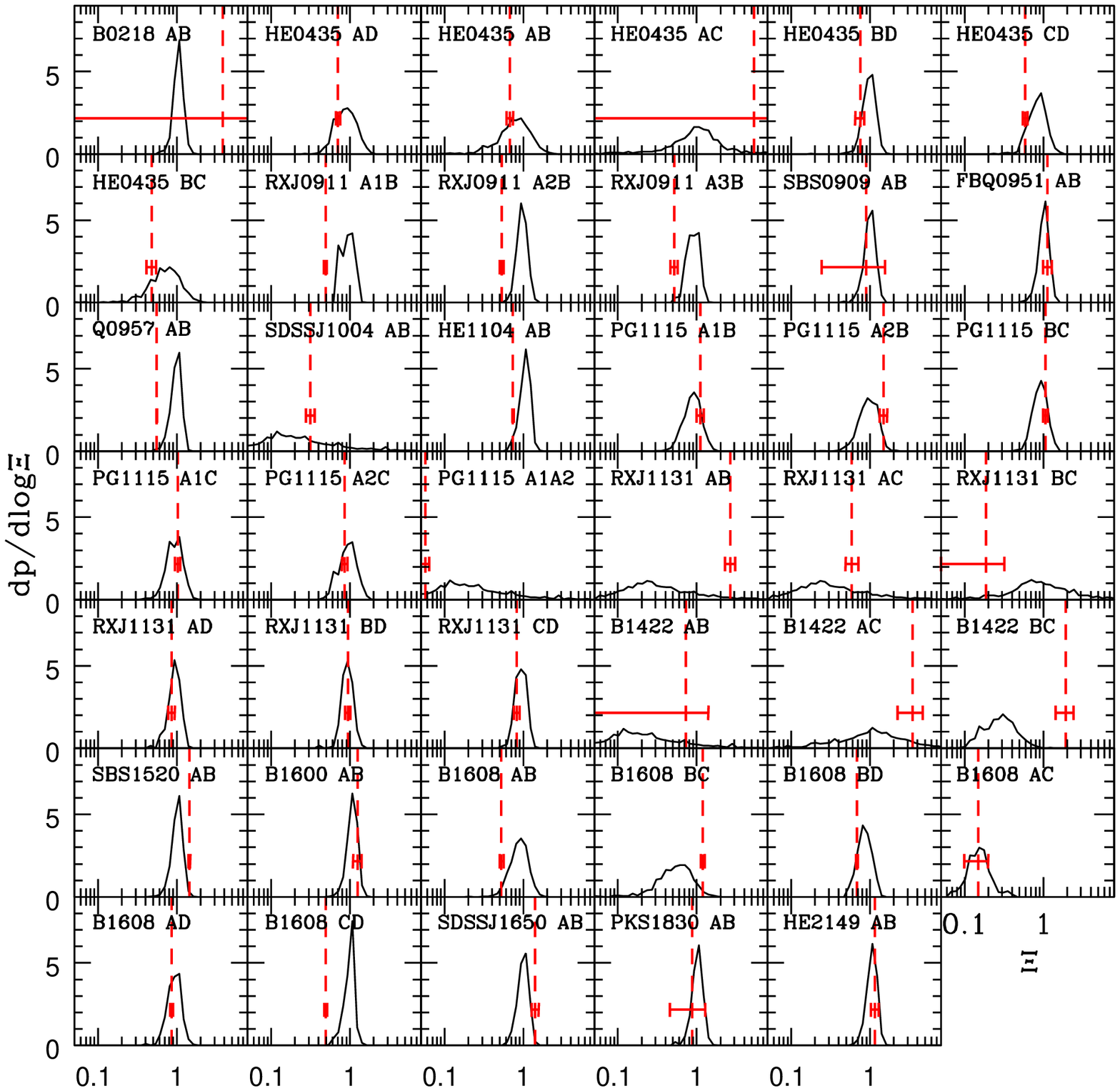}
\caption{The conditional probability distributions
  $p(\Xi|R_{ij},\theta_{ij})$ are compared with reduced time delays
  $\Xi$ computed from observed time delays of 17 published time delay
  quasars. Again, the Hubble constant of $h=0.73$ is assumed. 
  Solid line curves are in fact differential probability
  distribution . In each panel, we draw the differential probability
  distribution $dp/d\log\Xi$ computed from observed values of $R_{ij}$
  and $\theta_{ij}$ by a solid line. Reduced time delays from observed
  time delays are shown by vertical dashed lines plus error-bars. 
  See Table \ref{tab:xi} for the summary of observations. 
\label{fig:obspdf}}
\end{figure*}

\section{Comparison with Observed Time Delays}\label{sec:obs}

We now examine if the conditional distribution computed in \S
\ref{sec:pot} is consistent with the observed distribution of time
delays. Since we need to assume the Hubble constant in order to
convert observed time delays to reduce time delays $\Xi$, in this
section we adopt $h=0.73$ that was obtained from the combined analysis 
of the cosmic microwave background (CMB) anisotropy and clustering of
galaxies \citep{tegmark06}. We derive reduced time delays for the 17
published time delay quasars (41 image pairs), which is summarized in
Table \ref{tab:xi}. 

First, we compare these reduced time delays with the
conditional probability plotted in Figure \ref{fig:dist}. We show 
$p(\Xi|R_{ij})$ and $p(\Xi|\theta_{ij})$ with observed values
overplotted in Figure \ref{fig:dist_obs}. It appears that the data 
are roughly consistent with our probability distribution from
simulations. The large scatter of very symmetric lenses is also seen
for observed time delays. It is interesting that observed time delays
appear to exhibit the decline of time delays at small opening angles,
just as our theoretical model predicts. We may need more quasar time
delays to confirm this trend observationally. 

More directly, observed time delays should be compared with the
probability distribution for given $R_{ij}$ and $\theta_{ij}$, i.e.,
$p(\Xi|R_{ij},\theta_{ij})$. In Figure \ref{fig:obspdf} we compare 
observed $\Xi$ with the probability distribution of $\Xi$ expected
from corresponding image configuration of each image pair. We find that
the distributions agree with observed values on average, similarly as
Figure \ref{fig:dist_obs}. The reduced time delays of B0218+357 and
SBS0909+532 have larger errors because the positions of the lens galaxies
are poorly determined. The large error of HE0435$-$1223 AC comes from
small $r_j^2-r_i^2$, whereas that of B1422+231 is simply because of
the large time delay measurement errors. We comment that RXJ0911+0551
and Q0957+561 has significantly smaller $\Xi$ compared with our model
prediction. This is clearly because of the cluster convergence: The two
lenses lie in near the centers of clusters and therefore the observed
time delays are pushed down by the convergence coming from dark matter
in the clusters (see \S \ref{sec:hub}). The reduced time delays of
B1608+656 and SBS1520+530 are largely offset from the predicted
values, and this is probably because of satellite galaxies in the lens
systems that significantly affect the time delays. The large time
delays of RXJ1131$-$1231 AB and AC were also noted by
\citet{morgan07}: Our result indicates that the broadened theoretical
distributions due to small perturbations are enough to explain the
observed high values of the time delays. The large offsets of
B1422+231 and SDSS J1650+4251 may come from the large uncertainties of
measured time delays (see \S \ref{sec:hub}).

\section{Implications for $H_0$}\label{sec:hub}

In this section, we turn the problem around and constrain the Hubble
constant using the conditional probability distribution constructed from
the Monte-Carlo simulations. Although the current sample of observed
time delays summarized in Table \ref{tab:xi} is somewhat heterogeneous
and may not be appropriate for the statistical study, we do this to
demonstrate how we can constrain the Hubble constant from the
probability distribution. We basically take all lens systems listed in
Table \ref{tab:xi}, but adopt the following setup to reduce systematic
errors. 
\begin{itemize}
 \item We do not use the AB time delay of SDSS J1004+4112. It is a lens
       system caused by a massive cluster of galaxies,
       thus our input distribution, which is designed for galaxies, does
       not represent a fair distribution of lens potentials. Moreover,
       the center of the lens potential appears to be offset from the
       position of the brightest cluster galaxy \citep{oguri04a}, which
       makes it inaccurate to estimate important parameters such $\Xi$,
       $R_{ij}$, and $\theta_{ij}$.
 \item Both RXJ0911+0551 and Q0957+561 are known to reside in the cluster
       environment, thus they are significantly affected by the cluster
       convergence $\kappa_{\rm clu}$. Since the mass-sheet degeneracy
       says $\Xi\propto \Delta t\propto 1-\kappa_{\rm clu}$, we divide
       reduce time delays $\Xi$ for these two systems by $1-\kappa_{\rm
       clu}$ in order to {\it deconvolve} the effect of the cluster
       convergence. As the values of $\kappa_{\rm clu}$, we adopt 
       $\kappa_{\rm clu}=0.3\pm0.04$ for RXJ0911+0551 \citep{hjorth02}, and
       $\kappa_{\rm clu}=0.26\pm 0.08$ for Q0957+561 \citep{fischer97}. 
\end{itemize}
In summary, we use 16 lensed quasar systems (40 image pairs) to
constrain the Hubble constant. For each image pair, we compute the
likelihood as follows: 
\begin{equation}
 \mathcal{L}_p(h)=\int \frac{dp}{d\Xi}(\Xi|R_{ij,{\rm
  obs}},\theta_{ij,{\rm obs}})G(\Xi|\Xi_{\rm obs}(h))d\Xi,
\label{eq:like}
\end{equation}
where $R_{ij,{\rm obs}}$, $\theta_{ij,{\rm obs}}$, $\Xi_{\rm obs}$ are
those for this specific image pair listed in Table \ref{tab:xi}, and 
$G(\Xi|\Xi_{\rm obs}(h))$ indicates the Gaussian distribution with
median $\Xi=\Xi_{\rm obs}$. Note that calculating $\Xi_{\rm obs}$ from
observed time delays require the Hubble constant $h$, hence
$\mathcal{L}_p$ is a function of $h$. Then we compute the effective
chi-square by summing up the logarithm of the likelihoods: 
\begin{equation}
 \chi_{\rm eff}(h)=\sum_{\rm quasar}\frac{1}{n_p}\sum_{\rm
  pair}\left[-2\ln\mathcal{L}_p(h)\right]. 
\end{equation}
The first summation runs over lens systems, whereas the second summation
runs over image pairs for each lens system; the number of pairs for
each lens is denoted by $n_p$. Note that $n_p=1$ all double lens
systems, and a quadruple lens system should have $n_p \leq {}_4{\rm
  C}_2=6$ depending on how many time delays have been observed for the
lens system. The factor $1/n_p$ was introduced such that all lens
systems have equal weight on the effective chi-square irrespective of
the number of image pairs. We derive the best-fit value and its error
of $h$ by the standard way using a goodness-of-fit parameter
$\Delta\chi_{\rm eff}\equiv \chi_{\rm eff}-\chi_{\rm eff}({\rm min})$.  

\begin{figure}
\epsscale{.95}
\plotone{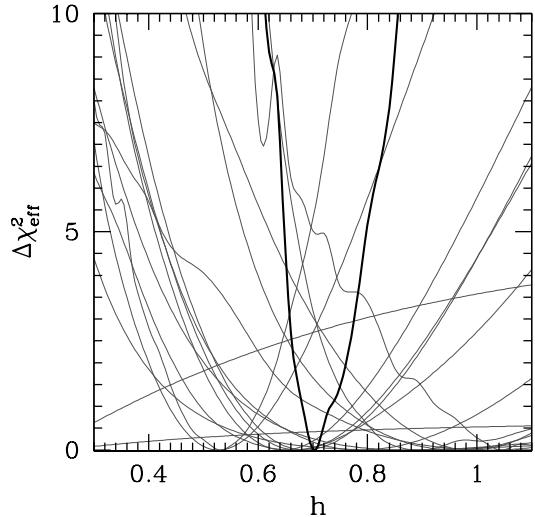}
\caption{Statistical constraint on the Hubble constant from 16 time
  delay quasars (40 image pairs). Thick solid line indicates
  goodness-of-fit parameter from all 16 lens systems plotted as a
  function of the Hubble constant $h$. The resulting Hubble constant
  is $h=0.70^{+0.03}_{-0.02}$ at 68\% confidence and  $h=0.70^{+0.09}_{-0.05}$
  at 95\% confidence. The Hubble constant estimated using jackknife
  resampling has a larger error, $h=0.70\pm0.06$ at 68\% confidence (see
  text for details). Thin solid lines show goodness-of-fit parameter
  for each lens system. 
\label{fig:hubble}}
\end{figure}

We show our result in Figure \ref{fig:hubble}. The Hubble constant
measured from the combination of all 16 lens systems is
$h=0.70^{+0.03}_{-0.02}$ at 68\% confidence and
$h=0.70^{+0.09}_{-0.05}$ at 95\% confidence. The obtained value is in
good agreement with other estimates, such as  the local distance
measurement using Cepheid calibration \citep{freedman01} and the CMB
anisotropy \citep{tegmark06,spergel07}. The constraint from each lens
system, which is plotted in Figure \ref{fig:hubble}, is summarized in
Table \ref{tab:hub}.

\begin{deluxetable}{lc}
\tablewidth{0pt}
\tabletypesize{\footnotesize}
\tablecaption{Hubble Constant from Each Lens System}
\tablewidth{0pt}
\tablehead{
 \colhead{Lens Name} & \colhead{$h$ ($1\sigma$ range)}}
\startdata
   B0218+357      & 0.21 (--)\\
   HE0435$-$1223  & 1.02 (0.70--1.39)\\
   RXJ0911+0551   & 0.96 (0.75--1.21)\\
   SBS0909+532    & 0.84 (0.47--)\\
   FBQ0951+2635   & 0.67 (0.56--0.81)\\
   Q0957+561      & 0.99 (0.82--1.17)\\
   HE1104$-$1805  & 1.04 (0.92--1.22)\\
   PG1115+080     & 0.66 (0.49--0.84)\\
   RXJ1131$-$1231 & 0.79 (0.59--1.03)\\
   B1422+231      & 0.16 (--0.36)\\
   SBS1520+530    & 0.53 (0.46--0.61)\\
   B1600+434      & 0.65 (0.54--0.77)\\
   B1608+656      & 0.89 (0.77--1.20)\\
   SDSS J1650+4251& 0.53 (0.44--0.63)\\
   PKS1830$-$211  & 0.88 (0.58--)\\
   HE2149$-$2745  & 0.69 (0.57--0.82)\\ \hline
   All            & 0.70 (0.68--0.73)
\enddata
\tablecomments{The Hubble constant and its error are estimated from
  the effective chi-square.}  
\label{tab:hub}
\end{deluxetable}

We also derive the Hubble constant using the jackknife resampling by
regarding each 16 lens system as a subsample. The result
$h=0.70\pm0.06$ at 68\% confidence has the same mean but larger error
than that estimated from the effective chi-square. There are several
possible source of this difference. One is the underestimate of the
width of the input distributions. In particular, many of the time delay
quasar systems has been claimed to be affected by lens galaxy environments 
\citep[e.g.,][]{morgan05,fassnacht06,momcheva06,williams06,auger07}, 
and thus our input strength of external shear might be somewhat
smaller than the true one (see also discussion in \S
\ref{sec:dis}). Another possible source is the non-Gaussianity of
measured time delays: In equation (\ref{eq:like}) we assumed the
Gaussian distribution for the measurement uncertainties of time
delays, but sometimes they are quite different from the Gaussian
distribution.\footnote{Among time delays listed Table \ref{tab:xi},
  those of SDSS J1650+4251 and B1422+231 could be significantly
  different from the true values (C. S. Kochanek, private
  communication). We perform the same analysis excluding these two
  systems and find the Hubble constant to be  $h=0.70^{+0.03}_{-0.04}$ at 
68\% confidence from the effective chi-square. Therefore our result is
not biased significantly by these systems.}  
We note that in our method we can in principle include non-Gaussianity
by just replacing $G(\Xi)$ in equation (\ref{eq:like}) with any
appropriate probability distributions, as long as we know such
distributions. 

\section{Discussions and Conclusion}\label{sec:dis}

In this paper, we have studied time delays between multiply imaged
quasars. Adopting the reduced time delay, which is a measure of how the
lens potential is complicated compared with the simple isothermal form,
we have explored the dependence of time delays on various complex
structure of lens potentials such as external perturbations,
non-isothermality, and substructures. The distribution of time delays
has been studied as a function of image configuration which we
characterize using two dimensionless quantities, the asymmetry and
opening angle of an image pair. We have pointed out that the
sensitivity on lens potentials is quite dependent on the image
configuration. For instance, more symmetric image pairs are more
affected by a small change of the lens potential. Image pairs with
smaller opening angles are also more sensitive to lens potentials. In
particular time delays of close image pairs are very sensitive to
higher-order external perturbations and substructures that are very
hard to be constrained from mass modeling even for best studied lens
systems. In addition, those pairs usually have the largest
relative uncertainties of time delay measurements. 
Therefore we conclude that it is quite difficult to extract
any information on the Hubble constant or the mass distribution from
close image pairs. It is interesting to note that perturbations
on lens potential not only introduce scatter around the mean but also
can systematically shift the distribution of time delays. One such
example is smaller time delays for smaller opening angle image pairs,
which is caused by external shear. 

We have performed Monte-Carlo simulations to derive a probability
distribution of reduced time delays for each image configuration.
Input distributions are determined from observational and theoretical
constraints. The distribution is weighted by the lensing cross section
and magnification bias, allowing a realistic estimate of time delay
distributions. The probability distribution was then compared with
observed time delays. We have shown that the distribution of time delays
computed from our simulations is in good agreement with observed time
delays. In particular, distributions of observed time delays also
exhibit strong dependence of image configuration in a consistent manner
with our theoretical expectations. The probability distribution can be
used to constrain the Hubble constant. We have found that 16 published
time delay quasars constrain it to be $h=0.70^{+0.03}_{-0.02}$ at 68\%
confidence using the effective chi-square or $h=0.70\pm0.06$ estimated
using jackknife resampling, consistent with other estimates.  

An important caveat is that our lensed quasar sample is quite
heterogeneous. In particular, it should be noted that current time delay
quasars (see Table \ref{tab:xi}) have significantly larger image
separations on average compared with the other quasar
lenses: The median image separation of time delay quasars listed
in Table \ref{tab:xi} is $1\farcs7$ (image separations before and
after the median are $1\farcs5$ and $2\farcs1$), whereas that of all
lensed quasars is $\sim 1\farcs4$. Quasar lenses with larger image
separations are more likely to lie in dense environments because both
the image separation and biased cross section are boosted by
surrounding dark matter \citep{oguri05b,oguri06a}, thus the Hubble
constant inferred from those lenses are more affected by the
environmental convergence. Indeed, the association of group/cluster
has been reported for more than half of the time delay quasars 
\citep[e.g.,][]{morgan05,fassnacht06,momcheva06,williams06,auger07}.
In addition, our input distributions of external perturbations may be
underestimated for these large separation lenses. To estimate this
systematic effect, we exclude four lensed quasars with image
separation larger than $3''$ and repeat the analysis done in \S
\ref{sec:hub}. The resulting Hubble constant $h=0.67^{+0.04}_{-0.03}$
at 68\% confidence from the effective chi-square is consistent with our full
result, thus we conclude that the effect of lens galaxy environments
is not so drastic here. However, to minimize the systematic effect, in
the future we should apply our statistical method to well-defined
samples of lensed quasars such as the CLASS \citep{myers03,browne03},
SQLS \citep{oguri06b}, and those obtained in future lens surveys. 
  
Another source of the systematic effect is the uncertainty of our
input distribution of lens potentials. Among others, the most
important systematic error comes from the uncertainty of the mean
value of the slope of the radial profile, $\alpha$. While
$\alpha=1$ (isothermal) for the mean appears to be a reasonable
choice, direct studies of lens galaxies
\citep[e.g.,][]{treu04,rusin05,hamana07} indicate that the error on
the mean $\alpha$ could be as large as 0.1. The derived Hubble
constant depends on the slope as $h\propto 2-\alpha$, therefore the
change of the mean $\alpha$ systematically shifts the best-fit value.
The scaling relation suggests that the 0.1 error of the mean results
in 10\% error on $h$, indicating that the systematic error may be
even larger than the statistical error. Another important systematic
effect is caused by external convergence from lens galaxy
environments. Since its effect on the Hubble constant is straightforward
\citep[e.g.,][]{keeton04}, one can estimate the effect of external
convergence rather easily even without including the distribution in the
simulation. From the result of \citet{oguri05b}, we expect that the
posterior distribution of external convergence with lensing bias taken
into account is roughly $\kappa_{\rm ext}\sim 0.03\pm0.03$ unless the
image separation is too large. This, combined with the fact that the
Hubble constant scales as $h\propto 1-\kappa_{\rm ext}$, suggest that
the effect of external convergence is not so dominant here (note that
external convergence was already taken into account for two extreme
lenses, Q0957+561 and RXJ0911+0551). By including
rough estimates of these two systematic errors, we obtain
$h=0.68\pm0.06({\rm stat.})\pm0.08({\rm syst.})$, indicating the
importance of reducing the systematic error.

Since the Hubble constant is now determined fairly well by other
methods, time delays are sometimes used to study mass distributions
of lens objects. Our statistical technique offers a new method to study
the lens mass distribution. By comparing the probability distributions
for different input distributions of lens potentials (e.g., different
median slopes of the primary lens galaxy), one can infer which input model is
most plausible. Unlike previous statistical studies 
\citep[e.g.,][]{oguri02}, this new method allows us to include various
complexity of lens potentials relatively easily, particularly the
non-spherically symmetric nature of lens potentials.  We note that our
input distribution of lens potentials was designed for galaxies, but it
is straightforward to modify it to that of lensing by other populations,
e.g., lensing by a cluster of galaxies.

In summary, our new statistical approach is invaluable for the study
of both cosmological parameters and structure of lens potentials.  We 
believe its importance grows more and more in the era of large-scale
synoptic surveys such as LSST and SNAP: Quasar lens candidates are
easily recognized in these synoptic surveys by making use of strong
time variability of quasars \citep{pindor05,kochanek06c}. Strong
lensing of distant supernovae offers additional interesting
opportunity to apply our statistical technique. The statistical
analysis is essential to make efficient use of the large homogeneous
samples of strong lenses provided by these surveys.

\acknowledgments
I thank Roger Blandford, Phil Marshall, Ted Baltz, Chris Kochanek, 
Xinyu Dai, Greg Dobler, Neal Jackson, Leon Koopmans, Chung-Pei Ma,
and Issha Kayo for discussions and comments. I am 
grateful to an anonymous referee for his/her many useful suggestions.   
This work was supported in part by the Department of Energy contract
DE-AC02-76SF00515.

\end{document}